\documentclass[fleqn,12pt]{wlscirep}
\usepackage[utf8]{inputenc}
\usepackage[T1]{fontenc}
\usepackage{floatrow} 
\usepackage{caption}
\usepackage{helvet}  
\usepackage{float} 
\title{Stiffness and Buckling Behavior of Woven Columns}

\makeatletter
\renewcommand{\@makecaption}[2]{%
  {\fontfamily{phv}\selectfont\footnotesize\textbf{#1.} #2\par}%
}
\makeatother 

\author[1]{Jaimie Krankel}
\author[2]{Guowei Wayne Tu}
\author[1,2,*]{Evgueni T. Filipov}
\affil[1]{Deployable and Reconfigurable Structures Laboratory, Department of Mechanical Engineering, University of Michigan, Ann Arbor, MI, USA}
\affil[2]{Deployable and Reconfigurable Structures Laboratory, Department of Civil Engineering, University of Michigan, Ann Arbor, MI, USA}
\affil[*]{E-mail: filipov@umich.edu}


\begin{abstract}
Woven shell structures are beneficial for applications requiring lightweight, damage resilience, and design tunability, such as in wearable devices, soft robotics, and aerospace systems. A fundamental component of woven structures is the woven column. While the mechanical properties of a woven column can be determined using sophisticated finite element (FE) simulations, these FE models are computationally expensive and do not explain the underlying mechanics behind scaling relationships. In this work, we derive  purely analytical models for the buckling load and stiffness of woven columns, and discuss the criteria that lead to different buckling modes of the woven columns. The simulated results based on our models closely match experimental data across various weave design parameters. This work advances our understanding of the mechanics of woven systems and serves as a baseline for the design of next-generation hierarchical structures and materials. 
\end{abstract}
\begin{document}

\flushbottom
\maketitle

\thispagestyle{empty}

\section{Introduction}
\begin{figure*}[!htb] 
\centering
\begin{floatrow}[1]
\ffigbox[\FBwidth]{\caption{\textbf{Overview of the woven columns and their buckling behaviors which are explored in this work}. \textbf{a.} Fabrication of a plain woven column, which is a fundamental unit for 3D woven shell structures. We assemble the vertical and horizontal weavers (left) into a woven column (right) using a plain weave pattern. \textbf{b.} A comparison shows that a woven column (top) does not experience permanent damage after buckling, whereas a continuous column (bottom) made with the same amount of material experiences plastic deformation and fracture. \textbf{c.} Buckling modes of twelve different woven columns, where localized out-of-plane deformations of weavers contribute to buckling of the columns. The buckling pattern is dependent on the parameters of the vertical and horizontal weavers. Scale bars are 5 cm.}\label{fig:1}}%
{\includegraphics[width=18cm]{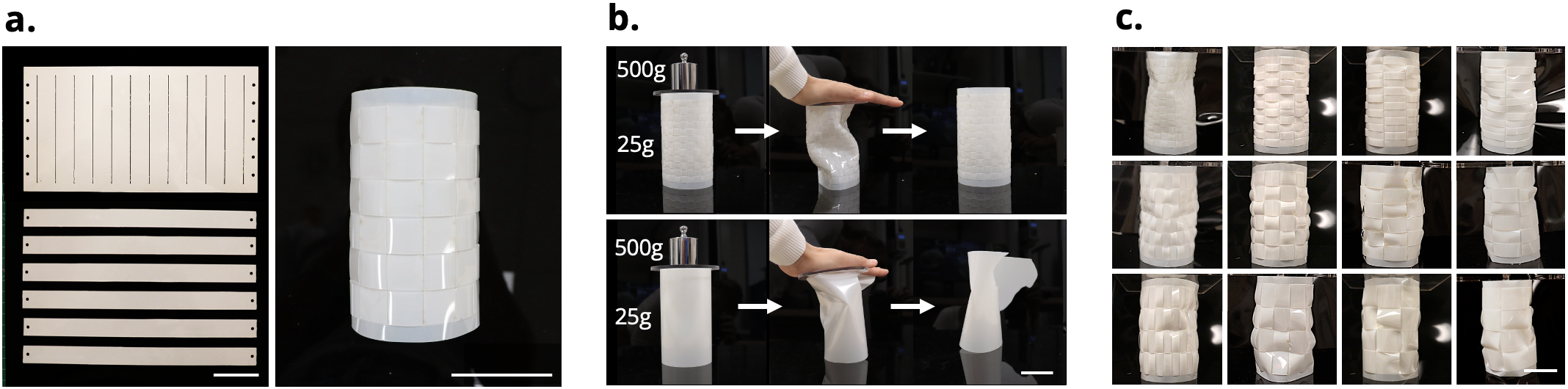}}
\end{floatrow}
\end{figure*}

The long-standing craft of weaving has gained traction in modern engineering due to its mechanical strength, damage tolerance, and lightweight properties. These intrinsic benefits of weaving make it suitable for smart material devices, soft robotics, aerospace applications, and more \cite{wayne,tu2025engineering,tu2024origami,JING2025113262,WANG2022115246}. The plain weave (Fig. \ref{fig:1}a), in which perpendicular weavers alternate over and under each other, is the most commonly implemented pattern. This simple weave pattern achieves tight interlocking of material, making for strong and durable structures \cite{ABUBAKAR2013575}. Previous work shows that plain woven 3D shell structures have much greater resilience \cite{wayne} yet they share similar stiffness to their continuous counterparts, as demonstrated in Fig. \ref{fig:1}b. 

Similar to the continuous cylindrical shell, the woven column is a fundamental unit of 3D woven shell structures. Cylindrical shell buckling has been widely studied, and there exist closed-form solutions for their behaviors\cite{hilburger2020buckling, hutchinson2018knockdowns, MA2023110330}. Woven column buckling behaviors, however, are more complicated due to local interactions between weavers (Fig. \ref{fig:1}c). Recent work on woven materials has focused on implementing finite element analysis techniques \cite{KUMAR2025107736, JUNIOR2024104274, NILAKANTAN20102300, Nguyen2018} and semi-analytical modeling \cite{LUO2021108503, Dabiryan2022, ElMessiry2018} to determine mechanical properties such as stiffness and buckling behaviors. These models require large computational power, and unlike an analytical model, they can only provide limited insight on the underlying mechanics of the system behavior. Moreover, the mechanical behavior of the cylindrical woven column unit has remained unexplored.

In this work, we derive purely analytical models for the stiffness and buckling forces of woven columns. Through physical experiments we explore the axial behavior of the column and verify the analytical models. We then classify global and local buckling modes of woven columns, and determine relationships between the buckling mode and the width of weavers. Our models and findings explain scaling laws where the stiffness and buckling load of woven columns change with the thickness and width of weavers. This work provides tools for choosing suitable weaver parameters for 3D woven column structures when they are to be designed for various potential engineering applications.

\section{Mechanics model derivation}\label{Sec:Modeling}
We use fundamental mechanics theory to derive models for the stiffness and critical buckling loads of a thin-walled, densely woven column. These models are based on geometric parameters and material properties of the vertical and horizontal weavers we use to fabricate the columns.

\subsection{Buckling of woven columns}
We assume that prior to buckling, the deformation of the woven columns is small and linear. Thus, we estimate that the critical buckling load of a woven column is the sum of the critical load of its vertical weavers and that of its horizontal weavers: $P_{cr, \ total} = P_{cr, \ h} + P_{cr, \ v}$. We first consider the critical load of vertical weavers $P_{cr, \ v}$. To simplify our model, we assume that the adjacent vertical weavers in a woven column initially buckle independently. Furthermore, by assuming a densely woven cylinder, each segment of a vertical weaver along its length also behaves independently. The load on a weaver with a small initial out-of-plane displacement can be approximated by the buckling force of an initially straight weaver since this force is minimally greater. Illustrated in Fig. \ref{fig:2}a, the buckling force of the combined vertical weavers is then governed by the buckling force of a singular segment. Based on the Euler buckling theory, $P_{cr, \ v}$ is dependent on the width of vertical weavers $w_v$, the thickness of vertical weavers $t_v$, and the width of horizontal weavers $w_h$ (which is also the length of each segment of a vertical weaver): \cite{hibbeler2015mechanics}
\begin{equation}\label{eqn:P_cr,v}
 P_{cr, \ v} = n_v\frac{\pi^2E\cdot w_vt_v^3}{w_h^2}, 
\end{equation}
where $n_v$ is the total number of vertical weavers and $E$ is the Young's modulus of the material.

Next, we consider the critical buckling load of the horizontal weavers $P_{cr, \ h}$. These weavers are assumed to be short cylindrical shells with axisymmetric sinusoidal initial perturbations. Based on previous research on thin-walled cylinders \cite{hilburger2020buckling}, the critical buckling load of a horizontal weaver is dependent on its normalized thickness $t_h/R$ (where $t_h$ is the thickness of horizontal weavers and $R$ is the radius of the woven column) and the surface area of the horizontal weaver $A_f$. We use Koiter's knockdown factor $\beta$ to account for the influence of the initial sinusoidal perturbation on the buckling load: \cite{hutchinson2018knockdowns}
\begin{equation}\label{eqn:P_cr,h}
P_{cr, \ h} = \beta\cdot\frac{E(t_h/R)}{\sqrt{3(1-\nu^2)}}\cdot A_f,
\end{equation}
where $\nu$ is the Poisson's ratio. Axisymmetric sinusoidal perturbations greatly reduce the buckling force compared to a perfect cylindrical shell. To account for this influence, Koiter's knockdown factor $\beta$ is solved implicitly using material parameters and the imperfection perturbation size $\delta_h$ shown in Fig. \ref{fig:2}b \cite{koiter1963knockdowns}:
 \begin{equation}\label{eqn:beta}
 \beta\cdot \Big(\frac{\delta_h}{t_h}\Big) = \Big(\frac{4}{27}(1-\nu^2)\Big)^{1/2}(1-\beta)^2.
 \end{equation}
The perturbation term $\delta_h$ in Eq. \ref{eqn:beta} is calculated by approximating the horizontal weaver as a polygonal section with $n_v$ sides where the length of each side is $w_v$. Then $\delta_h$ is the average distance of this polygon from a circle of equal circumference. Equivalently, we take half the distance between its maximum and minimum radii as in Fig. \ref{fig:2}b to calculate the $\delta_h$:
\begin{equation}\label{eqn:delta_h}
 \delta_h = \frac{1}{2}\bigg(\frac{w_v}{2\sin(\frac{\pi}{n_v})}+\frac{w_v}{2\tan(\frac{\pi}{n_v})}\bigg).
\end{equation} 

We compute the buckling force of the horizontal weavers using Eqs. (\ref{eqn:P_cr,h}), (\ref{eqn:beta}), and (\ref{eqn:delta_h}). We then add it to the buckling force of the vertical weavers Eq. (\ref{eqn:P_cr,v}) to obtain the buckling force of the entire woven column. 

\subsection{Stiffness of woven columns}
\begin{figure*}[!htb] 
\centering
\begin{floatrow}[1]
\ffigbox[\FBwidth]{\caption{\textbf{Derivation of the stiffness and the buckling load of a woven column.} \textbf{a.} Schematic of the vertical weavers. Each segment of a vertical weaver is modeled as an independent column undergoing buckling. To derive stiffness, the inner force and bending moment are analyzed at locations of maximum curvature. \textbf{b.} Schematic of the horizontal weavers. Horizontal weavers are modeled as axisymmetric sinusoidally perturbed cylindrical shells stacked on top of one another. The horizontal cross-section is approximated as a polygon with $n_v$ sides. Given a cross-section, the perturbation $\delta_h$ is derived by averaging $R_{max}$ and $R_{min}$. The contact area $A$ for deriving the horizontal stiffness is taken as the total contact area between adjacent horizontal weavers.}\label{fig:2}}%
{\includegraphics[width=12cm]{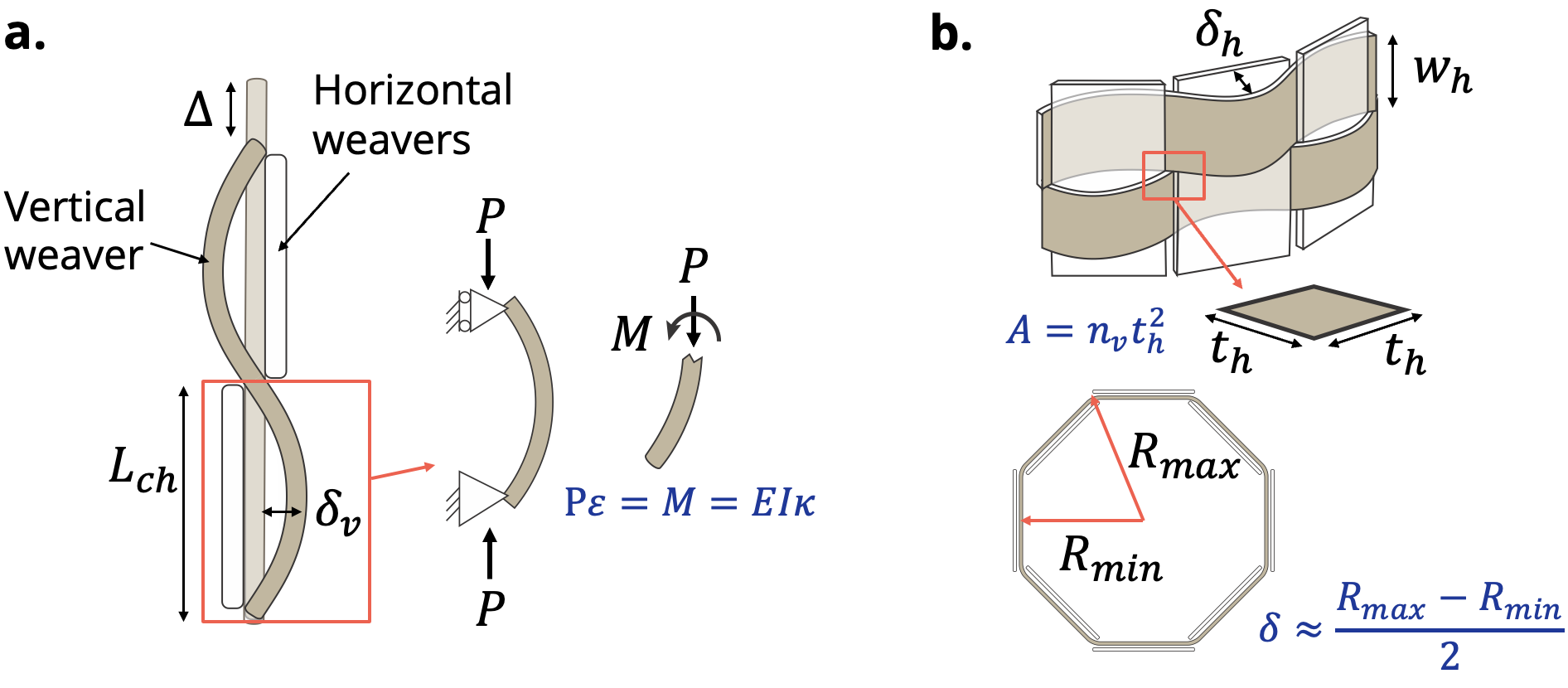}}
\end{floatrow}
\end{figure*}

Considering the linear deformation where the displacement is infinitesimally small, we assume that the stiffness of a woven column is the sum of the stiffness of its horizontal weavers $k_h$ and the stiffness of its vertical weavers $k_v$: $k_{total} = k_{h} + k_{v}$. We first consider the vertical weaver stiffness $k_v$. We assume that the vertical weavers deform sinusoidally as they are compressed, and that bending deformations store significantly more energy than axial deformations. All vertical weavers act in parallel and resist bending deformations. The slender vertical weavers mainly bend about the favored axis of bending (the horizontal axis), but they also bend about the less favored axis of bending (the vertical axis). Therefore, the vertical weavers take on a curved cross-section due to the nature of weaving, and this pre-curvature increases their stiffness significantly by a factor $\alpha$ \cite{Pini2016, Taffetani2019, PhysRevLett.113.214301}. We balance forces and moments at the location of the greatest perturbation (as in Fig. \ref{fig:2}a) to obtain a dependence of the axial load of the weaver $P_v$ on flexural rigidity $EI$, number of vertical weavers $n_v$, out-of-plane curvature of the weaver $\kappa$, and linear displacement perturbation $\delta_v$. Note that $\kappa$ and $\delta_v$ are both functions of the axial displacement $\Delta$. We calculate the stiffness as the derivative of axial load with respect to axial displacement:
\begin{equation}\label{eqn:k_v1}
{k_v} = \alpha \cdot \frac{n_v}{n_h} \cdot \frac{d}{d\Delta} \bigg(EI\frac{\kappa(\Delta)}{\delta_v(\Delta)} \bigg)
\end{equation}
The factor $\alpha$ accounts for the significant effect of curvature induced stiffness \cite{Pini2016, Taffetani2019, PhysRevLett.113.214301}, which increases with vertical weaver width. We apply Pini's equation \cite{Pini2016} to the vertical weaver segments, using the radius of the woven column $R$:
\begin{equation}\label{eqn:alpha}
 \alpha=\frac{w_v^4}{60t_v^2}\big(\delta_v(\frac{n_h\pi}{L-\Delta})^2-\frac{\nu}{R}\big)^2+1
\end{equation}

Assuming the out-of-plane displacement $\delta_v$ is small and that each vertical weaver deforms sinusoidally with $n_h$ half-waves, we determine the maximum curvature $\kappa$ of the vertical weavers using the curvature formula \cite{colley2018vector}. At the location where the maximum curvature occurs, the curvature reduces to
\begin{equation}\label{eqn:curvature}
\kappa = \delta_v \Big(\frac{n_h\pi}{L-\Delta}\Big)^2. 
\end{equation}
Because $\kappa \propto \delta_v$ in Eq. (\ref{eqn:curvature}), the derivative in Eq. (\ref{eqn:k_v1}) simplifies without further calculation of $\delta_v$. Therefore, we combine Eqs. (\ref{eqn:k_v1}), (\ref{eqn:alpha}), and (\ref{eqn:curvature}) into
\begin{equation}\label{eqn:k_v}
k_v = \Big(\frac{w_v^4}{60t_v^2}\big(\delta_v(\frac{n_h\pi}{L-\Delta})^2-\frac{\nu}{R}\big)^2+1\Big)\cdot 2\frac{n_v}{n_h}EI\frac{(n_h\pi)^2}{(L-\Delta)^3}.
\end{equation}

Horizontal weavers act in series and deform axially in response to loading. The contact area $A$ is the area between vertical weavers where adjacent horizontal weavers make contact, as shown in Fig. \ref{fig:2}b. Thus, the contribution of the horizontal weavers to stiffness is
\begin{equation}\label{eqn:k_h}
k_{h} = \frac{En_vt_h^2}{n_hw_h}.
\end{equation}

We sum Eqs. (\ref{eqn:k_v}) and (\ref{eqn:k_h}) to obtain the overall stiffness of a woven column.

\section{Fabrication and experimental methods}
We conducted a parametric study and validated our models through physical experimentation on woven cylindrical column samples. We constructed samples varying: (1) vertical weaver thickness $t_v$, (2) horizontal weaver thickness $t_h$, (3) vertical weaver width $w_v$, (4) horizontal weaver width $w_h$, and (5) sample height $h$. We construct a base woven column with $t_v = t_h = 0.191$ mm, $w_v = w_h = 20$ mm, and $h=140$ mm. We then conduct five experiments, each keeping one parameter constant and varying the other four. 

\begin{figure*}[!htb] 
\centering
\begin{floatrow}[1]
\ffigbox[\FBwidth]{\caption{\textbf{Fabrication of woven columns and the test setup}. \textbf{a.} Assembly of a column where vertical weavers are connected at the top and bottom. This design maintains consistent spacing in samples, but can only be used for testing in which the thickness of vertical weavers remains constant. Scale bars are 4 cm. \textbf{b.} Assembly of a column where vertical weavers are mechanically joined using split pins at the top and bottom of horizontal weavers. This design maintains consistent end effects, and is only used in a few of our tests where vertical thickness varies. Scale bars are 4 cm. \textbf{c.} Test setup using Mark-10\textsuperscript{\textregistered} ESM 1500 single-column tabletop testing system. \textbf{d.} Schematic for plate-to-plate compression loading of woven columns. Vertical and horizontal weavers are both assumed to be deformed with sinusoidal deflections. \textbf{e.} A typical force--displacement curve obtained from an experiment with the buckling force and stiffness measured.}\label{fig:3}}%
{\includegraphics[width=15cm]{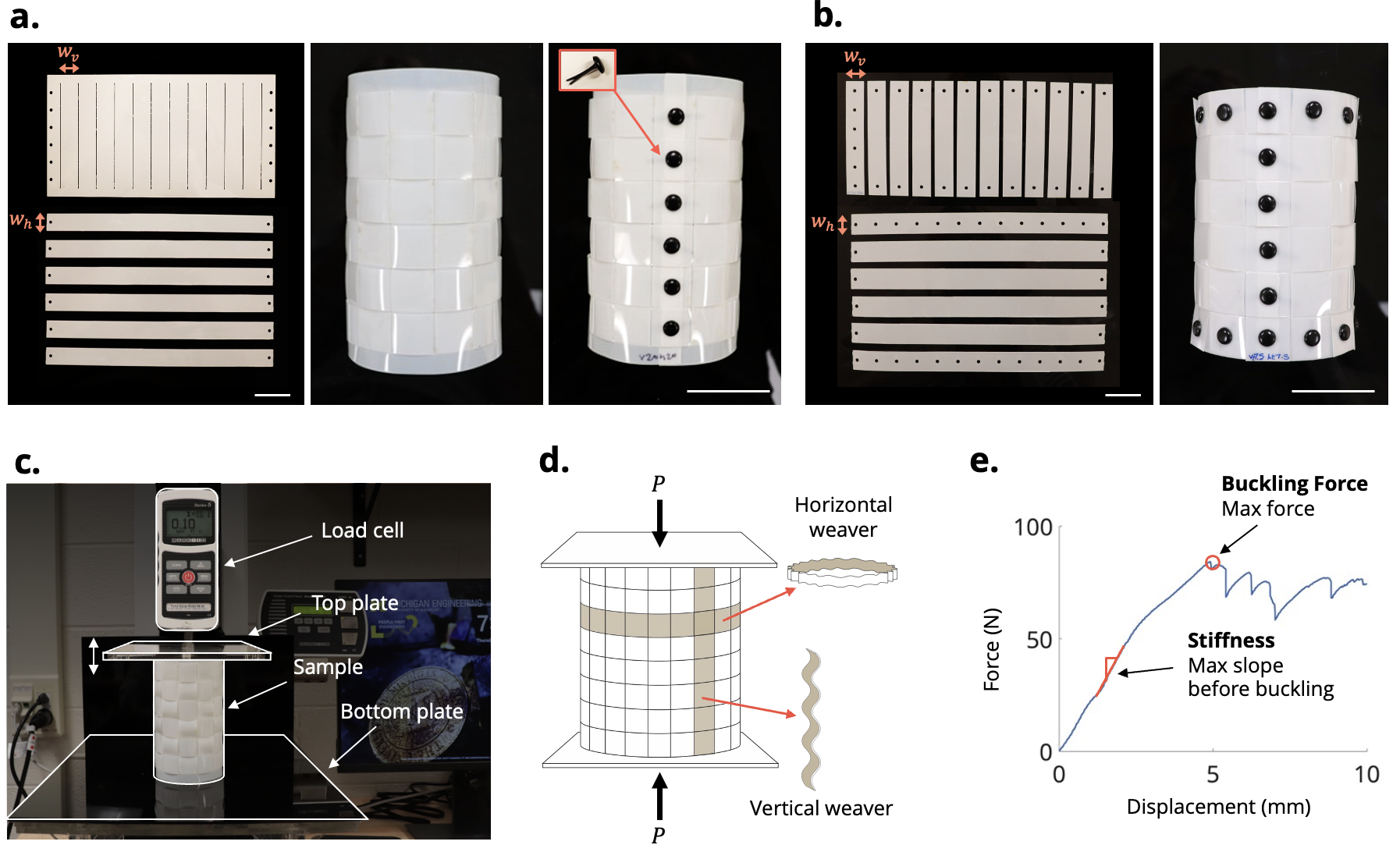}}
\end{floatrow}
\end{figure*}

Our samples are woven by hand from vertical and horizontal strips of Mylar\textsuperscript{\textregistered} polyester connected by a vertical seam of split pins (Fig. \ref{fig:3}a). For studies with consistent vertical weaver thickness, we cut the vertical weavers from a continuous Mylar\textsuperscript{\textregistered} sheet and leave a 10 mm connection at the top and bottom (Fig. \ref{fig:3}a). This connection at the two ends assists fabrication and facilitates consistency in spacing, but may cause varying end effects when the vertical weaver thickness varies. In our parametric study where the thickness of vertical weavers varies, we instead connect vertical weavers using two additional rows of split pins (Fig. \ref{fig:3}b). We maintain consistent tightness of the weaving across all samples by scaling the spacing between weavers proportionally to the thickness of weavers\cite{banerjee2014weaving, lord1982weaving}.

We obtained the experimental buckling force, stiffness, and qualitative buckling mode for each woven sample using plate-to-plate compression loading between two acrylic plates (Fig. \ref{fig:3}b, c, and d). The samples were compressed at a rate of 15 mm/min using a Mark-10\textsuperscript{\textregistered} ESM 1500 single-column tabletop testing system with a 250 N load cell. Force and displacement are recorded at a sampling rate of 20 Hz until a global maximum force is reached. Buckling force is taken as the peak load experienced and stiffness is taken by numerically differentiating the data to find the maximum instantaneous slope before the buckling force is reached (Fig. \ref{fig:3}e). To account for precision error in sample fabrication, we tested three identical samples of each variation and calculated the average values and error ranges. 

\section{Results}
Parametric studies were performed by varying horizontal weaver width $w_h$, vertical weaver width $w_v$, horizontal weaver material thickness $t_h$, vertical weaver material thickness $t_v$, and sample height $h$. We observed and classified the different buckling modes of woven columns, and we validated our mechanics models for the buckling force and stiffness of woven columns against the experimental data.

\subsection{Buckling mode classification}
\begin{figure*}[!htb] 
\centering
\begin{floatrow}[1]
\ffigbox[\FBwidth]{\caption{\textbf{Comparison of buckling modes}. \textbf{a.} Depending on the widths of horizontal and vertical weavers in a woven column, the buckling mode of the column switches between global, local, and combination modes. \textbf{b.} Typical force--displacement curve for a column that experiences global buckling into a diamond pattern, with no pre-buckling behavior observed. \textbf{c.} Typical force--displacement curve for a column that experiences local buckling, where pre-buckling is observed, and local force maxima occur before a peak force is obtained.}\label{fig:4}}%
{\includegraphics[width=15cm]{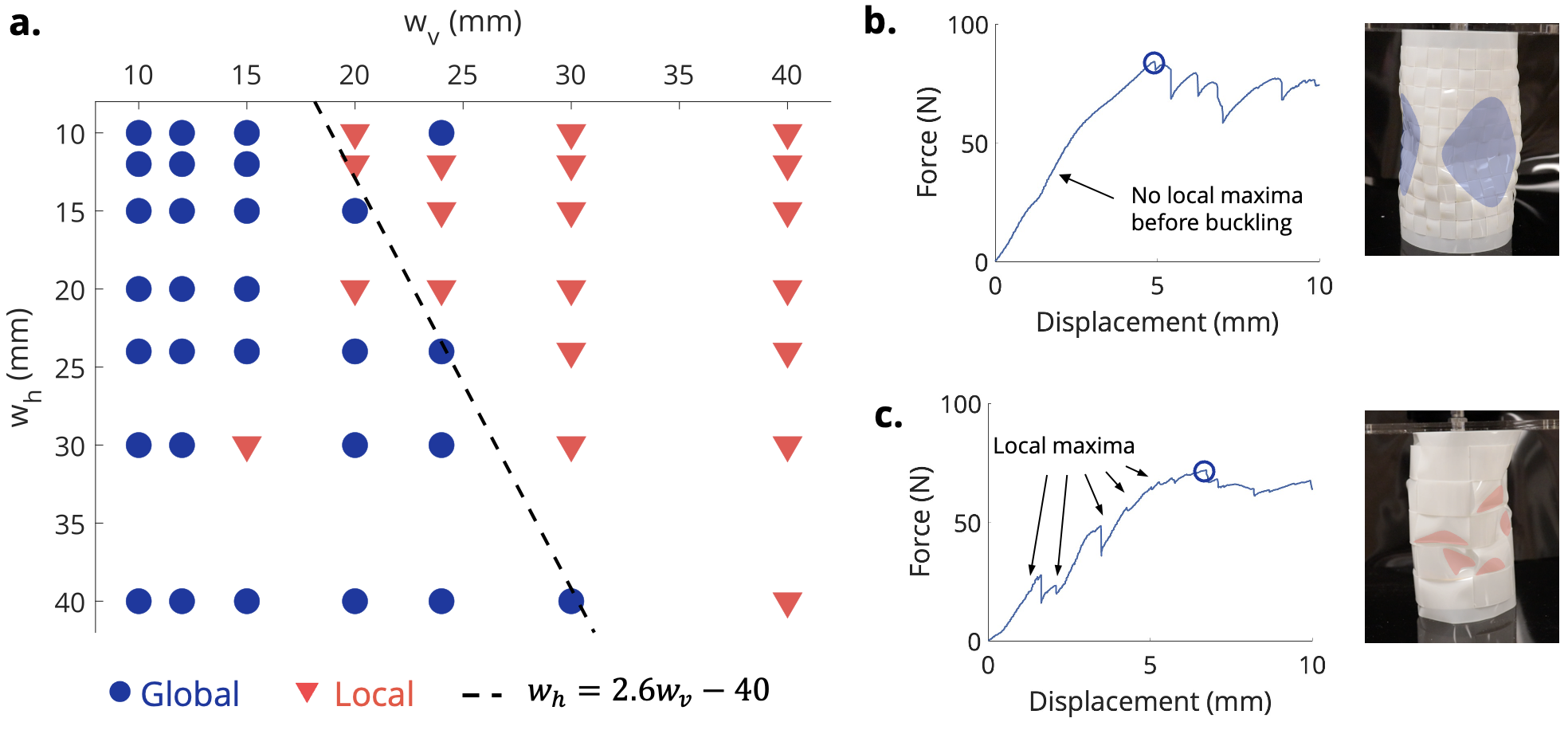}}
\end{floatrow}
\end{figure*}

In our parametric study, we vary the widths of vertical and horizontal weavers. We observed a strong correlation of buckling behavior with the weaver widths (Fig. \ref{fig:4}a). We define the buckling behavior of a column as local if it experiences any local maxima in its force-displacement curve before 95\% of the buckling force is reached. We define
 these local maxima as force values in which all forces within a $\pm$0.4 mm displacement range are less than 93\% of the local maximum value (Fig. \ref{fig:4}c). We define the buckling behavior of the column as global if it does not exhibit these local maxima before peak buckling is reached (Fig. \ref{fig:4}b). By using this criterion to distinguish local and global buckling modes, we can determine an approximate first-order boundary between local and global buckling regions by: $w_h=2.6w_v-40$. That is, if $2.6w_v-w_h>40$ then the column is more likely to buckle locally. Otherwise, it is more likely to buckle globally. 

For the columns that experience the global buckling mode, under large deformations the horizontal and vertical weavers tend to deform together, typically into a diamond pattern. Even for this global buckling scenario, we believe that local buckling of vertical and horizontal weavers remains the precipitating factor, consistent with our assumptions in Section \ref{Sec:Modeling}. Meanwhile, for local buckling modes, initial buckling can occur in multiple locations before the system reaches a peak load. We observe more global buckling in columns with wider horizontal weavers and narrower vertical weavers because these patterns allow the cylinder to buckle between horizontal boundaries. Meanwhile, when vertical weavers are wider and horizontal weavers are narrower, the additional curvature of the vertical weavers counteracts buckling across horizontal boundaries, which results in more complicated local behaviors. 

We observe that for columns experiencing global buckling, the force-displacement behavior before peak load is more predictable than for locally buckling columns. Therefore, applications requiring repeatable or predictable deformation such as in buckling-enabled soft robotics or mechanical computing may benefit from woven structures that undergo a global buckling mode. Our results suggest that the desired buckling modes for 3D woven structure can be specifically designed by adjusting the widths of the vertical and horizontal weavers.

\subsection{Scaling relationships of buckling load}
The buckling force of a woven column increases with both the thicknesses of its vertical and horizontal weavers, as validated in Fig. \ref{fig:5}a and b. Buckling force increases proportionally to the thickness of vertical weavers cubed ($P_{\mathrm{cr}}\propto t_v^3$), resulting from the relationship of vertical weaver buckling force to the second moment of inertia (Fig. \ref{fig:5}a). Buckling force increases with the thickness of horizontal weavers as a result of the horizontal weaver buckling force component (Fig. \ref{fig:5}b).

Shown in Fig. \ref{fig:5}d and e, our buckling force model is validated against variation in the widths of vertical and horizontal weavers. Increases in the width of vertical weavers $w_v$ cause greater initial sinusoidal imperfections in horizontal weavers, making the horizontal weavers and the entire column more prone to buckling (Fig. \ref{fig:5}d). When the width of horizontal weavers $w_h$ is varied, our idealized model predicts that for small widths of horizontal weavers, the vertical weavers will experience higher-order sinusoidal wave patterns and higher buckling forces (Fig. \ref{fig:5}e). In reality, inconsistencies in weaving allow for weak spots where the vertical weavers form uneven waves and the resulting buckling force is lower than predicted. These defects become more likely when horizontal weavers are less wide and vertical weavers have more segmental waves, so a realistically fabricated column will not experience substantial increases in buckling forces as the horizontal weaver widths become small (Fig. \ref{fig:5}e). This effect on buckling force is most apparent in our data when horizontal and vertical widths are both less than 15 mm. In contrast, when the width of horizontal weavers $w_h$ increases, the vertical weavers have few segmental waves and the buckling force approaches a constant, which is accurately captured by our theoretical model (Fig. \ref{fig:5}e).

When considering the column height, there is a negligible effect on the buckling force compared to other parameters (Fig. \ref{fig:5}c). Our model, similar to classical models for cylindrical shell buckling \cite{hutchinson2018knockdowns}, does not account for the height of the column, so it matches this trend well. This result further supports our initial assumption, where local buckling of the vertical and horizontal weavers, rather than global system buckling, drive the behavior of the system.

\begin{figure*}[!htb] 
\centering
\begin{floatrow}[1]
\ffigbox[\FBwidth]{\caption{\textbf{Influence of design parameters on buckling forces of woven columns}. Buckling force with respect to: \textbf{a.} Vertical weaver thickness, \textbf{b.} Horizontal weaver thickness, \textbf{c.} Column height, \textbf{d.} Vertical weaver width, \textbf{e.} Horizontal weaver width. \textbf{f.} Schematic of all design parameters. }\label{fig:5}}%
{\includegraphics[width=15cm]{figure5.pdf}}
\end{floatrow}
\end{figure*}

\subsection{Scaling relationships of stiffness}
\begin{figure*}[!htb] 
\centering
\begin{floatrow}[1]
\ffigbox[\FBwidth]{\caption{\textbf{Influence of design parameters on the stiffness of woven columns}. Stiffness with respect to: \textbf{a.} Vertical weaver thickness, \textbf{b.} Horizontal weaver thickness, \textbf{c.} Column height, \textbf{d.} Vertical weaver width, \textbf{e.} Horizontal weaver width. \textbf{f.} Schematic of all design parameters.}\label{fig:6}}%
{\includegraphics[width=15cm]{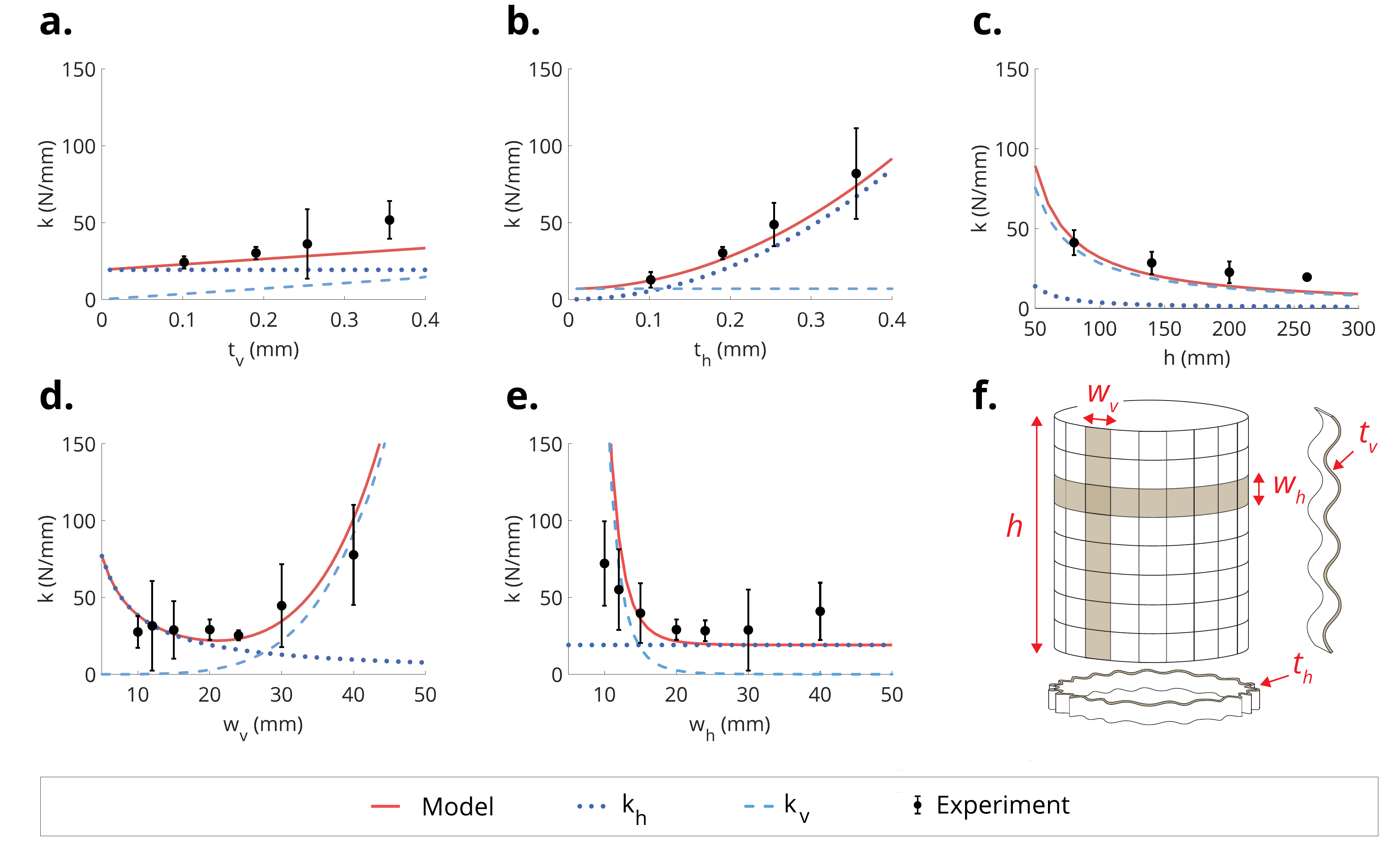}}
\end{floatrow}
\end{figure*}

The stiffness of the woven column increases with respect to both its vertical and horizontal weaver thicknesses, as validated in Fig. \ref{fig:6}a and b. By increasing vertical weaver thickness $t_v$, we increase the bending modulus of the vertical weavers by a factor of thickness cubed $t_v^3$, resulting in a stiffer structure (Fig. \ref{fig:6}a). By increasing the horizontal weaver thickness $t_h$, we increase the horizontal weaver stiffness proportionally to their thickness squared ($k\propto t_h^2$). This increase is because the contact area between adjacent horizontal weavers increases with the thickness squared (Fig. \ref{fig:6}b). 

Shown in Fig. \ref{fig:6}d and e, our stiffness model is validated against variation in vertical and horizontal weaver width. As the vertical weaver width $w_v$ increases, the vertical weavers exhibit more curvature about the less favored vertical axis of bending, resulting in additional curvature-induced stiffness from the vertical weavers (Fig. \ref{fig:6}d). As $w_v$ decreases, the contact area between horizontal weavers increases, resulting in increased stiffness due to the horizontal weaver component of stiffness. As the horizontal weaver width $w_h$ decreases, the vertical weavers have smaller periods with greater curvature relative to displacement, and thus higher stiffness (\ref{fig:6}e). In both cases, increasing the overall stiffness of the column can be achieved by having the vertical weavers contribute a larger portion of stiffness than the horizontal weavers.

When considering the height of the columns in Fig. \ref{fig:6}c, the stiffness increases for shorter columns because of the contribution of the horizontal weavers. The stiffness increases exponentially for shorter columns because when keeping $w_h$ constant, we have $h\propto n_h$ hence the relationship follows from $k_h\propto\frac{1}{n_h}$. 

\section{Conclusions}
We created analytical models that predict the buckling force and stiffness of cylindrical woven columns made from a dense planar weave. The models are based on the geometry of the columns and basic mechanics principles, which makes the models versatile and simple to use. The models have been validated with experimental data considering different parameters, including weaver widths, weaver thicknesses, and column height. Our models explain the underlying mechanics behind buckling force and stiffness trends in woven structures and serve as a closed-form tool for design. We determined that while both vertical and horizontal weavers contribute to the overall column buckling force and stiffness, in some cases one or the other will have a larger influence on tuning a specific property. When thickness of either horizontal or vertical weavers is increased, both the buckling capacity and the stiffness increase. When we increase the width of horizontal weavers, the buckling capacity remains constant and the stiffness decreases. When we increase the width of vertical weavers, the buckling capacity decreases while the stiffness increases. Our results can inform design for efficient material use by identifying which weaver parameters can be changed to achieve a desired structural property. 

We established predictive guidelines for the buckling mode of a column. We categorized a column's buckling mode as local if the column experiences local maxima in its force-displacement curve before buckling. A column's buckling mode is categorized as global if the column does not experience local maxima before buckling. These modes are largely dependent on horizontal and vertical weaver widths, and the transition between modes can be described by a linear relationship between horizontal and vertical weaver widths. Columns with larger vertical weaver widths and smaller horizontal weaver widths typically experience local buckling. Columns with smaller vertical weaver widths and larger horizontal weaver widths typically experience global buckling where the structure deforms into a diamond pattern. Our findings serve to guide the design of woven structures such that favorable behaviors and buckling modes can be achieved, and buckling load and stiffness can be analytically calculated. These hierarchical woven structures have potential for future engineering applications in soft intelligent robots, flexible metamaterials, customizable wearable devices, and more. 

\section*{Acknowledgements}
The authors acknowledge support from the Air Force Office of Scientific Research under award number FA9550-22-1-0321. The paper reflects the views and opinions of the authors, and not necessarily those of the funding entities. 

\bibliography{article}

@article{wayne,
  title={Corner topology makes woven baskets into stiff, yet resilient metamaterials},
  author={Tu, Guowei Wayne and Filipov, Evgueni T},
  journal={Physical Review Research},
  volume={7},
  number={3},
  pages={033193},
  year={2025},
  publisher={APS},
  doi = {https://doi.org/10.1103/9srl-9gsc},
  url = {https://link.aps.org/doi/10.1103/9srl-9gsc}
}

@article{tu2025engineering,
  author ="Tu, Guowei Wayne and Filipov, Evgueni T.",
title  ="Engineering snags for spatial curvature in weaves: fabrication{,} mechanics{,} and inverse design",
journal  ="Soft Matter",
year  ="2025",
volume  ="21",
issue  ="46",
pages  ="8793-8802",
publisher  ="The Royal Society of Chemistry",
doi  ="https://doi.org/10.1039/D5SM00813A",
url  ="http://dx.doi.org/10.1039/D5SM00813A",
}

@article{tu2024origami,
  title={Origami of multi-layered spaced sheets},
  author={Tu, Guowei Wayne and Filipov, Evgueni T},
  journal={Journal of the Mechanics and Physics of Solids},
  volume={190},
  pages={105730},
  year={2024},
  publisher={Elsevier},
  doi={https://doi.org/10.1016/j.jmps.2024.105730},
}

@book{hibbeler2015mechanics,
  title={Mechanics of Materials},
  author={Hibbeler, R.C.},
  year={2015},
  publisher={Pearson}
}

@book{banerjee2014weaving,
  title={Principles of fabric formation},
  author={PK Banerjee},
  year={2014},
  publisher={CRC Press}
}

@book{lord1982weaving,
  title={Weaving: Conversion of yarn to fabric},
  author={PR Lord},
  year={1982},
  publisher={Woodhead Publishing}
}

@article{JUNIOR2024104274,
title = {Concurrent multiscale modelling of woven fabrics: Using beam finite elements with contact at mesoscale},
journal = {Finite Elements in Analysis and Design},
volume = {242},
pages = {104274},
year = {2024},
issn = {0168-874X},
doi = {https://doi.org/10.1016/j.finel.2024.104274},
url = {https://www.sciencedirect.com/science/article/pii/S0168874X24001689},
author = {Celso Jaco Faccio Júnior and Vijay Nandurdikar and Alfredo Gay Neto and Ajay B. Harish},
}

@article{NILAKANTAN20102300,
title = {On the finite element analysis of woven fabric impact using multiscale modeling techniques},
journal = {International Journal of Solids and Structures},
volume = {47},
number = {17},
pages = {2300-2315},
year = {2010},
issn = {0020-7683},
doi = {https://doi.org/10.1016/j.ijsolstr.2010.04.029},
url = {https://www.sciencedirect.com/science/article/pii/S0020768310001654},
author = {Gaurav Nilakantan and Michael Keefe and Travis A. Bogetti and Rob Adkinson and John W. Gillespie},
}

@article{MA2023110330,
title = {Buckling analyses of thin-walled cylindrical shells subjected to multi-region localized axial compression: Experimental and numerical study},
journal = {Thin-Walled Structures},
volume = {183},
pages = {110330},
year = {2023},
issn = {0263-8231},
doi = {https://doi.org/10.1016/j.tws.2022.110330},
url = {https://www.sciencedirect.com/science/article/pii/S0263823122008825},
author = {He Ma and Peng Jiao and Hongfei Li and Zhi Cheng and Zhiping Chen},
}

@article{WANG2022115246,
title = {Many-objective design optimisation of a plain weave fabric composite},
journal = {Composite Structures},
volume = {285},
pages = {115246},
year = {2022},
issn = {0263-8223},
doi = {https://doi.org/10.1016/j.compstruct.2022.115246},
url = {https://www.sciencedirect.com/science/article/pii/S0263822322000587},
author = {Zhenzhou Wang and Adam Sobey},
}

@article{PhysRevLett.113.214301,
  title = {How a Curved Elastic Strip Opens},
  author = {Barois, Thomas and Tadrist, Lo\"{\i}c and Quilliet, Catherine and Forterre, Yo\"el},
  journal = {Phys. Rev. Lett.},
  volume = {113},
  issue = {21},
  pages = {214301},
  numpages = {5},
  year = {2014},
  month = {Nov},
  publisher = {American Physical Society},
  doi = {https://doi.org/10.1103/PhysRevLett.113.214301},
  url = {https://link.aps.org/doi/10.1103/PhysRevLett.113.214301}
}

@article{Taffetani2019,
  title = {Limitations of curvature-induced rigidity: How a curved strip buckles under gravity},
  author = {Taffetani, M. and Box, F. and Neveu, A. and Vella, D.},
  journal = {Europhysics Letters},
  volume = {127},
  number = {1},
  pages = {14001},
  year = {2019},
  publisher = {{EPLA}},
  doi = {https://doi.org/10.1209/0295-5075/127/14001}
}

@article{Pini2016,
  title = {How two-dimensional bending can extraordinarily stiffen thin sheets},
  author = {Pini, Valerio and Ruz, Jakub and Kosaka, Priscila and Maletinsky, Patrick and Meyer, Carsten and Esashi, Masayoshi and Blick, Robert H.},
  journal = {Scientific Reports},
  volume = {6},
  pages = {29627},
  year = {2016},
  publisher = {Nature Publishing Group},
  doi = {https://doi.org/10.1038/srep29627}
}

@article{LUO2021108503,
title = {Static, buckling, and free-vibration analysis of plain-woven composite plate with finite thickness using VAM-based equivalent model},
journal = {Thin-Walled Structures},
volume = {169},
pages = {108503},
year = {2021},
issn = {0263-8231},
doi = {https://doi.org/10.1016/j.tws.2021.108503},
url = {https://www.sciencedirect.com/science/article/pii/S0263823121006170},
author = {Dan Luo and Yifeng Zhong and Senbiao Xi and Zheng Shi},
}

@article{Dabiryan2022,
  author    = {H. Dabiryan and M. Jesri and H. R. Ovesy and Z. S. Mazloomi},
  title     = {Numerical and experimental study of buckling behavior of delaminated plate in glass woven fabric composite laminates},
  journal   = {Journal of Engineered Fibers and Fabrics},
  volume    = {17},
  year      = {2022},
  doi       = {https://doi.org/10.1177/15589250221091268},
  publisher = {SAGE Publications},
  url       = {https://doi.org/10.1177/15589250221091268}
}

@article{JING2025113262,
title = {Impact resistance of 3D woven fabrics and composites: A review},
journal = {Thin-Walled Structures},
volume = {213},
pages = {113262},
year = {2025},
issn = {0263-8231},
doi = {https://doi.org/10.1016/j.tws.2025.113262},
url = {https://www.sciencedirect.com/science/article/pii/S0263823125003568},
author = {Kunkun Jing and Suchao Xie and Yifan Zhang and Hui Zhou and Hongyu Yan},
}

@article{ABUBAKAR2013575,
title = {Optimization of elastic properties and weaving patterns of woven composites},
journal = {Composite Structures},
volume = {100},
pages = {575-591},
year = {2013},
issn = {0263-8223},
doi = {https://doi.org/10.1016/j.compstruct.2012.12.043},
url = {https://www.sciencedirect.com/science/article/pii/S0263822313000263},
author = {Ilyani Akmar {Abu Bakar} and Oliver Kramer and S. Bordas and Timon Rabczuk},
}

@techreport{hilburger2020buckling,
  author       = {Hilburger, Mark W.},
  title        = {Buckling of Thin-Walled Circular Cylinders},
  institution  = {NASA},
  type         = {NASA Special Publication},
  number       = {NASA SP-8007-2020/REV 2},
  address      = {Hampton, VA},
  year         = {2020},
  note         = {Revised version of NASA SP-8007 (1965)},
  url          = {https://ntrs.nasa.gov/api/citations/20205011530/downloads/20205011530%20Rev%202FINALa%201-2023.pdf}
}

@book{colley2018vector,
  author    = {Susan Jane Colley},
  title     = {Vector Calculus},
  edition   = {5th},
  year      = {2018},
  publisher = {Pearson},
  address   = {Boston},
  isbn      = {9780134766712}
}

@article{KUMAR2025107736,
title = {Modelling woven composites with shell elements: An application of second-order computational homogenisation},
journal = {Computers \& Structures},
volume = {312},
pages = {107736},
year = {2025},
issn = {0045-7949},
doi = {https://doi.org/10.1016/j.compstruc.2025.107736},
url = {https://www.sciencedirect.com/science/article/pii/S004579492500094X},
author = {Athira Anil Kumar and Aewis K.W. Hii and Stephen R. Hallett and Bassam El Said}
}

@article{ElMessiry2018,
  author       = {El Messiry, M. and El-Tarfawy, S.},
  title        = {Mechanical properties and buckling analysis of woven fabric},
  journal      = {Textile Research Journal},
  year         = {2018},
  volume       = {89},
  number       = {14},
  pages        = {2900--2918},
  doi          = {https://doi.org/10.1177/0040517518803777}
}

@article{Nguyen2018,
  author    = {Q. T. Nguyen and T. D. Nguyen and D. V. Dao and D. V. Le and T. T. Nguyen},
  title     = {A study on mechanical properties of 3D printing ABS plastic according to different printing orientations},
  journal   = {IOP Conference Series: Materials Science and Engineering},
  volume    = {459},
  number    = {1},
  year      = {2018},
  publisher = {IOP Publishing},
  doi       = {https://doi.org/10.1088/1757-899X/459/1/012082}
}

@article{hutchinson2018knockdowns,
  title     = {On Establishing Buckling Knockdowns for Imperfection-Sensitive Shell Structures},
  author    = {Gerasimidis, S. and Virot, E. and Hutchinson, J. W. and Rubinstein, S. M.},
  journal   = {Journal of Applied Mechanics},
  volume    = {85},
  number    = {9},
  year      = {2018},
  publisher = {ASME},
  doi       = {https://doi.org/10.1115/1.4040455},
}

@article{koiter1963knockdowns,
  title={The Effect of Axisymmetric Imperfections on the Buckling of Cylindrical Shells under Axial Compression},
  author={Koiter, W.T.},
  journal={Proc. K. Ned. Akad. Wet., B},
  volume={66},
  year={1963},
  pages={265--279},
  number  = {5}, 

}

\end{document}